\documentclass{epl}
\usepackage{graphicx}
\usepackage{dcolumn}
\usepackage{bm}
\usepackage{epsfig,psfrag,amsmath,amssymb,float}
\usepackage{color}
\newcommand{\vv}[1]{{\bf #1}}

\def\cha#1{{#1 }}

\title{
Role of vertex corrections in the spin-fluctuation mediated pairing
mechanism
}
\shorttitle{
Vertex corrections and spin fluctuations
}
\author{Z.~B.~Huang\inst{1,3}, W.~Hanke\inst{1}, and E.~Arrigoni\inst{1,2}}
\institute{\inst{1}Institut f\"ur Theoretische Physik und Astrophysik, 
Universit\"at W\"urzburg,
am Hubland, 97074 W\"urzburg, Germany\\
\inst{2} Institut f\"ur Theoretische Physik - Computational Physics, 
Technische Universit\"at Graz, Petersgasse 16, A-8010 Graz, Austria\\
\inst{3} Department of Physics, Hubei University, Wuhan 430062, China
}


\begin{document}

\maketitle

\begin{abstract}
We study numerically and partly diagrammatically
the renormalization of the electron-spin interaction or 
vertex in a two-dimensional one-band Hubbard model
with spin-fluctuation momentum transfer ${\vv q}=(\pi,\pi)$.
We find that the renormalized electron-spin vertex decreases 
quite generally with decreasing temperature at all doping densities.
As a combination of two concurring effects, i.e. the decrease
of the vertex and the increase of the spin susceptibility,
the effective pairing interaction increases with lowering 
temperature in the intermediate-correlation regime, but 
{\it flattens off} in the strong-correlation regime.
Our findings indicate that in the high-T$_c$ cuprates the pairing 
mediated by antiferromagnetic spin fluctuations is substantially 
reduced due to vertex corrections.
\end{abstract}

\section{Introduction}
The spin-fluctuation mediated interaction between charge carriers is 
a corner stone in a variety of theoretical and experimental 
issues~\cite{scalapino,IFT98},
which are presently in the center of focus of the high-T$_c$ 
superconductivity research: it has been proposed that a spin-1 
resonance mode, which is prominent in the magnetic response measured
by neutron scattering, is responsible for significant features 
(``kink") seen in other spectroscopies such as 
photoemission~\cite{lanzara,norman,abanov} 
and optical 
absorption, which are sensitive to the charge dynamics, and even 
that this mode is the boson responsible for ``mediating" the 
superconducting pairing~\cite{carbotte}. 
This also has led to a variety of 
counter-arguments based on the small spectral weight of
the resonance mode and (assumed weak)
coupling to electron-hole pairs~\cite{kee}. 
However, neither the original proposals for the resonant 
spin-fluctuation mode scenarios nor the counter-arguments take 
into account the effects of vertex corrections. 
\cha{
The vertex
function $\gamma(\vv k,\vv q)=g_{\vv k\vv q}/g_{\vv k\vv q}^{0}$,
which gives the ratio of the coupling $g_{\vv k\vv q}$ 
of the boson ($\vv q = (\pi,\pi)$ magnon or resonance mode) with 
a dressed quasiparticle 
 (which includes
vertex corrections and also quasiparticle renormalization)
to the bare vertex
$g_{\vv k\vv q}^{0}$, 
}
is expected to be substantially 
renormalized by strong electronic correlations.
 This has recently 
been demonstrated for the electron-phonon (el-ph) vertex~\cite{zbhuang} 
by using numerically accurate quantum-Monte-Carlo (QMC)
techniques~\cite{BSS81}: whereas
at weak and intermediate Coulomb (Hubbard $U$) interactions, screening
is the dominant correlation effect suppressing the el-ph coupling,
at larger $U$ values irreducible vertex corrections become dominant
and give rise to an unexpected and substantial increase in the
forward direction. 

The need for clarifying the role of vertex corrections in the 
high-T$_c$ schemes involving the exchange of 
spin-fluctuations also derives
from a rather dramatic finding by Schrieffer~\cite{schrieffer}:
he argued that the spin-fluctuation vertex is strongly suppressed
in the long-ranged antiferromagnetic (AF) state or, in the 
paramagnetic state but with strong AF precursors, i.e.
\begin{equation}
\gamma(\vv k,\vv q)\propto [(\vv q - \vv Q)^{2}+
\frac{1}{\xi^{2}}]^{\frac{1}{2}}
\sim \frac{1}{[\chi(\vv q,\omega=0)]^{\frac{1}{2}}},
\label{gamma_sc}
\end{equation}
where $\xi$ 
stands for the AF correlation length and $\chi(\vv q,\omega=0)$
is the spin susceptibility. 
This finding is further confirmed by 
Chubukov {\it et al.}~\cite{chub}.
The vanishing of $\gamma$ at $\vv q = \vv Q \equiv (\pi,\pi)$
in the long-ranged AF state is a consequence of
the Adler principle: in the (AF)
ordered state, the magnons are Goldstone bosons and, therefore,
they always decouple from other degrees of freedom~\cite{chub}.
This then leads to a
strong renormalization of the pairing interaction $V$ in the 
single spin-fluctuation exchange approximation:
\begin{equation}
V(\vv k,\vv q)\propto |\gamma(\vv k,\vv q)|^{2}
\chi(\vv q),
\label{pair}
\end{equation}
which instead of diverging tends to a finite
positive constant as $\vv q$ approaches $Q$
and $\xi\rightarrow\infty$. On the other hand, for the appearance
of d-wave superconductivity one would need a substantial enhancement
of the interaction near $\vv Q$. There remains the question of 
how large the AF correlation length $\xi$ must be
for this reduction effect to be substantial. Furthermore, 
the theoretical argument uses the ``frozen in" electron-spin model,
where the holes move in an unaffected (by the presence of holes) 
AF background.

\begin{figure}
\centering
\epsfig{file=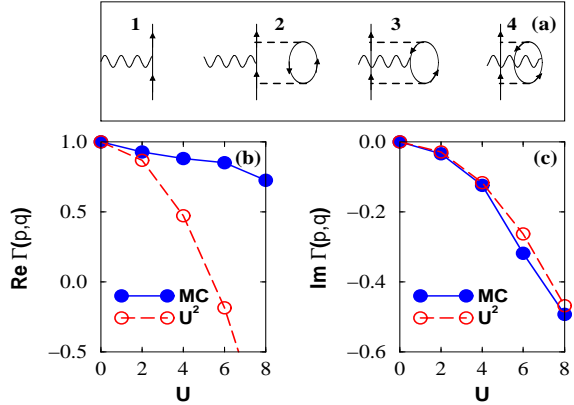,height=7.5cm,width=5.5cm,angle=270}
\caption{
(a) Low-order Feynman diagrams for the el-sp vertex
$\Gamma(p,q)$. The thin solid lines are the non-interacting Green's
functions and the dashed lines represent the Hubbard interaction $U$.
The wavy lines stand for the spin-fluctuation fields.
Real (b) and imaginary (c) parts of $\Gamma(p, q)$ vs $U$ at
$\delta=0.12$ and $\beta=2$. The solid circles are Monte Carlo results
and the open symbols show the perturbation theory contributions shown
in \ref{Diagram}(a).}
\label{Diagram}
\end{figure}

Using QMC techniques~\cite{BSS81}, 
we aim at clarifying the role of 
electronic correlations on the effective electron-spin fluctuation
(el-sp) coupling in a two-dimensional one-band Hubbard model.
By comparing QMC results at a typical low doping ($\sim 10\%$)
with diagrammatic ($\sim U^{2}$) calculations,
we infer that the ``weak correlation" regime, i.e. where 
$\gamma\sim 1$, is confined
to rather small $U$ ($U\leq 2t$, $t$: hopping) values 
(see Fig.~\ref{Diagram}). In this regime, and also
in the intermediate correlation regime ($U\sim 4t$) where the bare
vertex is already substantially suppressed ($\sim 50\%$), the strong 
increase of the susceptibility $\chi$ with decreasing
temperature dominates the effective
pairing potential $V$ in Eq.~(\ref{pair}).
Thus, $V(\vv p,\vv q =(\pi,\pi))$ with $\vv p$ close to the 
Fermi surface increases for decreasing temperatures. 
However, this standard behavior 
even qualitatively changes in the 
physically relevant strong correlation ($U\geq 8t$) regime: 
here, we do find in the underdoped regime that the vertex 
$\gamma$ is so substantially reduced at low enough temperatures
($T\leq J=4t^{2}/U$) that this reduction over-compensates the 
simultaneous increase of the spin susceptibility. Thus, 
at temperatures below the magnetic scale $J$, the vertex reduction
introduces a substantial suppression of the effective pairing potential.
At a first glance, this finding displays a similarity to
Schrieffer's observation (see Eq.~\ref{pair}). 
We will argue, however, that 
in our calculation the AF precursor has not developed and
that our findings are not consistent with a picture of
hole fermions embedded in a long-range AF background, but rather with
a short-range ``real-space" picture, valid for strong correlations.
Here, the spin fluctuation couples to the quasiparticle which 
forms a ``spin-bag", i.e. a hole dressed by short-range AF 
fluctuations.

\section{Model and numerical approach}
Our starting point is the one-band Hubbard model,
\begin{eqnarray}
\label{ham}
 H = -t \sum_{\langle ij \rangle,\sigma}
     (c_{i\sigma}^\dagger c_{j\sigma}^{\,}
     +c_{j\sigma}^\dagger c_{i\sigma}^{\,})
     + U \sum_i n_{i\uparrow}n_{i\downarrow},
\end{eqnarray}
The operators $c_{i\sigma}^\dagger$ and $c_{i\sigma}^{\,}$  as usual create 
and destroy an electron with spin $\sigma$ at site $i$, respectively and 
the sum $\langle ij\rangle$ is over nearest-neighbor lattice sites.
Here, $U$ is the onsite
Coulomb interaction and we will choose the nearest-neighbor hopping $t$ as
the unit of energy.

\begin{figure}
\centering
\epsfig{file=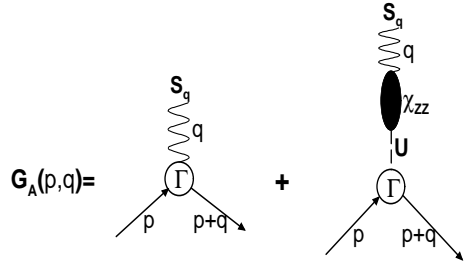,height=6.cm,width=3.6cm,angle=270}
\caption{Diagrammatic representation of $G_A(p,q)$
within linear response to $S_{\vv q}$. The thick solid lines
represent dressed single-particle Green's functions of
the Hubbard model. The wavy line denotes the external
perturbation in Eq.~(\ref{el-sp}). The dashed line represent
the Hubbard interaction $U$ and the closed ellipse stands for
the longitudinal spin susceptibility $\chi_{zz}(q)$.}
\label{svertex}
\end{figure}

In our simulations, we have used the linear-response 
technique as in Ref.~\cite{zbhuang} in order to extract the el-sp vertex.
In this method, one formally adds to Eq.~\eqref{ham} the interaction
with a momentum- and (imaginary) time-dependent spin-fluctuation field
in the $z$-direction
$S_{ \vv q} e^{-i q_0 \tau}$~\cite{pq} in the form~\cite{fint}
\begin{equation}
\label{el-sp}
H_{el-sp}= \sum_{\vv k \vv q\sigma} g_{\vv k\vv q}^{0}
\sigma c_{\vv k+\vv q\sigma}^{\dagger}c_{\vv k\sigma} \
S_{ \vv q} \ e^{-i q_0 \tau}\;,
\end{equation}
where $g_{\vv k\vv q}^{0}$ is the bare el-sp coupling (equal to
the Hubbard $U$ in the one-band Hubbard model).
One then considers the
``anomalous'' single-particle propagator in the presence of this
perturbation defined as~\cite{pq}
\begin{equation}
\label{gq}
G_{A}(p,q)\equiv 
- \int_{0}^{\beta}d\tau\ e^{i(p_{0}+q_{0}) \tau}
 \langle T_{\tau}c_{\vv p+\vv q\sigma}(\tau)c_{\vv p\sigma}^{\dagger}
(0)\rangle_{H+H_{el-sp}},
\end{equation}
Here $\langle\rangle_{H+H_{el-sp}}$ is the Green's function
evaluated with the Hamiltonian $H+H_{el-sp}$.
Diagrammatically, $G_A(p, q)$ has the structure shown in
Fig.~\ref{svertex} so that the el-sp vertex $\Gamma(p,q)$
can be expressed quite generally in terms of $G_A$ and
of the single-particle Green's function $G(p)$ in
the form
\begin{equation}
\Gamma (p, q) = \lim_{S_{\vv q}\to 0}\frac{1}{g_{kq}^{0}} 
\frac{1}{S_{\vv q}}
\frac{1}{1+U\, \chi_{zz}(q)}\frac{ G_A (p, q)}{G(p+q)\, G (p)} \;,
\label{five}
\end{equation}
with $\chi_{zz}(q)$ the longitudinal spin susceptibility.
Due to the limit procedure in Eq.~\ref{five},
it is sufficient to calculate the leading 
linear response of $G_A$ to $H_{\rm el-sp}$, which is given by
\begin{eqnarray}
G_A(p,q) = S_{\vv q} \
 \int_{0}^{\beta} d\tau e^{i(p_{0}+q_{0})\tau}
\int_{0}^{\beta} d\tau^{'} e^{-i q_{0} \tau'}
\sum_{\vv k\sigma^\prime} g_{\vv k\vv q}^{0}
\times \nonumber\\
\langle T_{\tau}\sigma^\prime c_{\vv k+\vv q\sigma^\prime}^{\dagger}
(\tau'+0^{+})c_{\vv k\sigma^\prime}(\tau')
 c_{\vv p+\vv q\sigma}(\tau)c_{\vv p\sigma}^{\dagger}(0)\rangle_{H},
\label{four}
\end{eqnarray}
where $0^+$ is a positive infinitesimal. Notice that $S_{\vv q}$
cancels in Eq.~\ref{five}.
The two-particle Green's function in Eq.~\eqref{four} is evaluated
with respect to the pure Hubbard Hamiltonian  (Eq.~\eqref{ham}).

\cha{
The effective vertex $\gamma$ describing scattering processes between
quasiparticles and spin waves is obtained by taking into account the
wave-function renormalisation
$Z(p)$:
}
\begin{equation}
\gamma(p, q) = \frac{\Gamma(p, q)}{\sqrt{Z(p)\, Z(p+q)}}\;.
\label{nine}
\end{equation}
\cha{
Numerically, $Z$ is obtained as
$Z(p)={\rm Im}[1/G(p)]/p_0$~\cite{zbhuang,pq}.
}
In terms of $\gamma(p,q)$ and of the spin susceptibility $\chi_{zz}(q)$,
the effective pairing interaction $V$ under the 
exchange of a single spin wave is expressed in the form
\begin{eqnarray}
V(p,q)&=&|\gamma(p,q)|^{2}U^{2}\chi_{zz}(q)\nonumber\\
&=&[|{\rm Re}\gamma(p,q)|^{2}+|{\rm Im}\gamma(p,q)|^{2}]
U^{2}\chi_{zz}(q),
\label{totalp}
\end{eqnarray}
with
\begin{eqnarray}
\chi_{zz} (q) & = &\frac{1}{2}\int^\beta_0 d\tau\ e^{-i\ q_0 \tau}
\ \left\langle T_\tau m_{\vv q}^{z} (\tau) m_{-\vv q}^{z} (0)
\right\rangle,
\nonumber\\
\noalign{\hbox{and}}
m_{\vv q}^{z} & = & {1\over\sqrt{N}}\sum_{\vv k\sigma}
\sigma c_{\vv {k+q}\sigma}^{\dagger}c_{\vv k\sigma},
\label{ten}
\end{eqnarray}
where $V(p,q)$ contains the contributions from both the real 
and imaginary parts of the vertex $\gamma$. 
\cha{
Notice that due to the renormalisation in Eq.~\ref{nine}, 
$V(p,q)$ describes the effective interaction {\it between
  quasiparticles}. 
}

Our numerical Monte Carlo simulations were performed on an 
$8 \times 8$ lattice at different doping densities and 
different temperatures~\cite{pq}.
In the high-T$_c$ cuprates, the charge carriers near the $(\pi,0)$ 
region are strongly affected by antiferromagnetic spin fluctuations. 
Therefore, we will examine the particular scattering process in which
the incoming electron and spin fluctuation carry momenta 
${\vv p}=(-\pi,\,0)$ and ${\vv q}=(\pi,\,\pi)$, respectively.
Within our $\vv p$-points mesh, these points lie sufficiently 
close to the Fermi surface.

\section{Results}
The diagrams contributing to the vertex $\Gamma$ up to order $U^2$ 
are displayed in Fig.~\ref{Diagram}(a).
Figs.~\ref{Diagram}(b) and \ref{Diagram}(c) show the comparison
of QMC calculations with perturbation theory. As one can see, 
the perturbative calculations are in good agreement with Monte 
Carlo simulations only for $U\lesssim 2$.
\begin{figure}
\centering
\epsfig{file=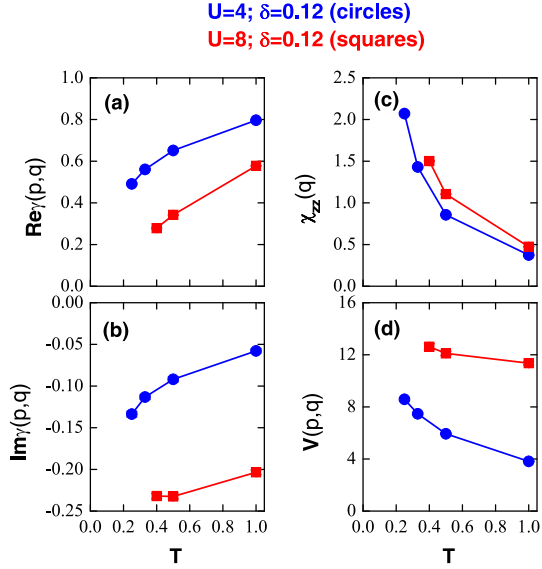,height=8cm,angle=0}
\caption{
Real (a) and imaginary (b) parts of the effective el-sp
vertex $\gamma(p,q)$ 
vs $T$ at $U=4$ and $U=8$ for the doping density
$\delta=0.12$. Figs.~(c) and (d) show the $T$-dependence of the spin
susceptibility $\chi_{zz}(q)$ and the effective pairing interactions
\cha{
between quasiparticles $V(p,q)$.
}
}
\label{Gam_Vsp_T}
\end{figure}

Monte Carlo results for $\gamma(p, q)$, $\chi_{zz}(q)$, and $V(p,q)$ 
are displayed in Figs.~\ref{Gam_Vsp_T}(a)-\ref{Gam_Vsp_T}(d).
We notice that in the underdoped regime ($\delta=0.12$)
for both intermediate correlation ($U=4$) and strong 
correlation ($U=8$), $\gamma$ is strongly renormalized 
at characteristic temperatures (below $T\approx J=0.5$ for $U=8$).
There is a clear temperature trend observable in Re$\gamma$, which
tends to go to zero (or to a very small value), at least for $U=8$.
At our lowest accessible $T$'s, Im$\gamma(p,q)$ is small
at weak and intermediate correlations, but for $U=8$ it can become
larger. Our numerical results clearly show that, whereas in the 
weak and intermediate correlation regimes the strong enhancement
of the spin susceptibility dominates the low-$T$ behavior of $V$,
which then strongly increases for decreasing $T$,
the behavior is different in the strong correlation ($U=8$) regime. 
Here, the pairing potential $V$ only displays a mild increase or a
saturation at low T's. Nevertheless, even at the lowest temperatures 
accessible at $U=8$, our conclusion that vertex corrections are crucial 
is already quite clear: there is an
order of magnitude reduction [$\sim O(10)$] in the effective pairing
interaction $V$ (Fig.~\ref{Gam_Vsp_T}(d)) compared with the RPA (with
full $\chi_{zz}$) result. 
Our finding at $U=4$ is in good agreement with the work of Bulut 
{\it et al.}~\cite{bulut}, which shows that the 
effective particle-particle interaction created by the Hubbard $U$ 
increases with lowering temperature and can reach large values.

\begin{figure}
\centering
\epsfig{file=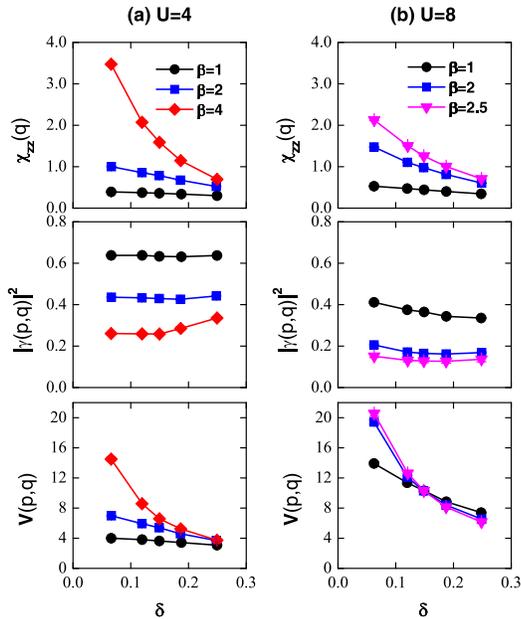,width=7cm,angle=0}
\caption{
Spin susceptibility $\chi_{zz}(q)$, the renormalization
factor $|\gamma(p,q)|^2$  (i. e. the reduction of $V$ with
respect to the RPA result), and the effective pairing interaction
$V(p,q)$ as a function of doping density $\delta$ for (a) $U=4$
and (b) $U=8$. The value of the inverse temperature $\beta$ is indicated
by the shape of the symbol.}
\label{Veff_gam2_chi}
\end{figure}

Figure~\ref{Veff_gam2_chi} gives for both the intermediate ($U=4$) 
and strong correlation ($U=8$) cases the doping dependence of 
the susceptibility $\chi_{zz}(q)$, of the renormalization factor 
$|\gamma(p,q)|^2$, and of the effective pairing interaction $V(p,q)$. 
Again, we note the competing influences of the temperature on $\chi$
and $\gamma$: $\chi$ increases both as a function of lowering doping
and temperature; this happens in both $U=4$ and $U=8$ cases. On the
other hand, the effective el-sp vertex $\gamma$ decreases as a function
of lowering temperature (as we have seen in 
Fig.~\ref{Gam_Vsp_T}), and is almost doping independent.
As already mentioned, at intermediate correlations, the $T$-dependence
of $\chi$ still dominates the pairing interaction, resulting in an increase
of the effective pairing upon reducing $T$, especially for low dopings. 
But even here,
at $U=4$, vertex corrections reduce $V$ substantially, as can be seen 
in the $|\gamma|^2$ value, which is reduced compared from the bare value
$1$ down to about $\sim 0.25$. This vertex influence on the 
effective pairing interaction is most dramatic at $U=8$. Here 
$|\gamma|^2$ reduces $V$ by a factor of $8$ at the lowest 
accessible $T$'s and dopings $\delta$. 

\section{Discussion and conclusions}
How can we understand the rather dramatic role of vertex corrections
in the strongly-correlated underdoped regime, 
where we have a short-ranged (of order Cu-Cu distance) 
correlation length $\xi$~\cite{groeber}
and not Schrieffer's
situation, where holes move in an unaffected (by the charge carriers)
AF ($\xi\rightarrow \infty$) background?
The important point is that
in our calculation, a strong vertex suppression is obtained {\it even
though the AF precursor has not developed}. 
What our QMC calculations 
demonstrate, is that below the characteristic temperature $T\sim J$,
vertex corrections become so large that in the physically relevant 
strong-correlation ($U=8$) regime, an order of magnitude reduction in
the pairing interaction results. 
This reduction happens more or less independent of doping. 
This doping  independence 
is at variance with Schrieffer's result (\ref{gamma_sc}), 
where of course $\xi$ depends crucially on doping.
Thus, our results suggest the following picture:
It is known from our QMC work on the single-particle spectral 
function $A({\vv k},\omega)$~\cite{groeber} that 
below $T\sim J$, a ``band" of width $O(J)$ forms, where
``spin-bag"-like quasiparticles propagate coherently. This happens 
again more or less independent of doping, i.e. from the underdoped
insulator ($\xi\rightarrow\infty$) up to optimal doping 
($\xi\approx$Cu-Cu distance). It can also be shown on the basis 
of exact diagonalization that it is the ``spin-bag" quasiparticle
and not the incoherent ``lower Hubbard band" background which 
couples most effectively to the perturbing spin potential $H_{el-sp}$ 
in Eq.~(\ref{el-sp})~\cite{eder}. 
Therefore, as soon as the quasiparticle
with its spin dressing has been formed (for low enough $T$'s and 
large enough $U$'s) the scattering amplitude or vertex is more or
less independent of doping (see Fig.~\ref{Veff_gam2_chi}).
This picture is clearly different from the doping dependent 
``screening effect" implicit in Eq.~(\ref{gamma_sc}).
\cha{
Our results were
obtained for a model with
nearest-neighbor hopping only. For a nonvanishing
next-nearest-neighbor hopping $t'/t<0$, spin fluctuations get
damped. In this case, calculations within the spin-fermion model show
that vertex corrections are reduced~\cite{chub}. 
 We carried our QMC
calculations for finite $t'/t=-0.25$,  showing that indeed  
vertex corrections are smaller 
 than for $t'/t=0$~\cite{details}, although the difference is not as pronounced as in
 Ref.~\cite{chub}.
}

In summary, based on quantum Monte Carlo simulations, we have studied
the renormalization of the el-sp interaction or vertex in the one-band
Hubbard model. We found that the renormalized el-sp vertex decreases
quite generally with decreasing temperature. On the other hand,
the temperature dependence of the effective pairing interaction is
rather different in the intermediate- and strong-correlation regimes:
It increases with lowering temperature in the intermediate-correlation
regime, but flattens off in the strong-correlation regime.

\acknowledgments
We would like to thank Prof. D.J.~Scalapino for useful discussions.
The W\"urzburg group acknowledges support by the DFG under 
Grant No.~DFG-Forschergruppe 538, by a Heisenberg Grant (AR 324/3-1), 
by the Bavaria California Technology Center (BaCaTeC), and the KONWHIR 
project CUHE. The calculations were carried out at the 
high-performance computing centers LRZ (M\"unchen) and
HLRS (Stuttgart).


\begin{thebibliography}{99}
\bibitem{pq} In our convention, unbolded variables denote both
Matsubara frequency and momentum, i.e., $p=(p_0,{\vv p})$ and
$q=(q_0,{\vv q})$. We have set the frequencies to their minimum 
values, i.e., $p_0=\pi T$ for fermions and $q_0=0$ for bosons.
\bibitem{fint}
The imaginary-time-dependent perturbation \eqref{el-sp}
has a well-defined meaning within a functional-integral formulation,
where it leads to the linear-response result \eqref{four}.
\end{thebibliography}
\end{document}